# Managing HIV/AIDS in Malawi

Brian G. Williams

South African Centre for Epidemiological Modelling and Analysis (SACEMA), Stellenbosch, South Africa
Correspondence to BrianGerardWilliams@gmail.com

**Abstract**

The epidemic of HIV in Malawi started earlier than in most of southern Africa and at its peak 15% of all adults were infected with HIV. Malawi is a low-income country and the cost of putting all HIV-positive people in Malawi onto ART, expressed as a percentage of the gross domestic product, is the highest in the world.[1] In spite of the scale of the epidemic and the cost of dealing with it, Malawi has made great progress and the effectiveness, availability and greatly reduced cost of potent anti-retroviral therapy (ART) make it possible to contemplate ending the epidemic of HIV/AIDS in Malawi. Here we consider what would have happened without ART, the *No ART* counterfactual, the impact on the epidemic if the current level of roll-out of ART is maintained, the *Current Programme*, and the likely impact if coverage is further expanded making treatment available to everyone who is eligible under the 2013 guidelines of the World Health Organization, starting in 2015 and reaching full coverage by 2020, the *Expanded Programme*.

The *Current Programme* has had a substantial impact on the epidemic of HIV and the number of people dying of AIDS. The *Expanded Programme* has the potential to avert more infections, save more lives and, most importantly, end the epidemic. The annual cost of managing HIV will increase from about US$132 million in 2014 to about US$155 million in 2020 but will fall after that. If the *Expanded Programme* is implemented several key areas must be addressed. Testing services will need to be expanded and supported by mass testing campaigns, so as to diagnose people with HIV and enrol them in treatment and care as early as possible. A regular and uninterrupted supply of drugs will have to be assured. The quantity and quality of existing health staff will need to be strengthened. Community health workers will need to be mobilized and trained to encourage people to be tested and accept treatment, to monitor progress and to support people on treatment; this in turn will help to reduce stigma and discrimination, loss to follow up of people diagnosed with HIV, and improve adherence for those on treatment.

## Introduction

The epidemic of HIV in Malawi started early. In 1987 HIV testing was done on workers on the South African gold mines in which correlations were made between their HIV status and their country of origin.[4] The men in the study came from South Africa, Malawi, Lesotho, Mozambique, Swaziland and Botswana but notably not from Rhodesia, now Zimbabwe. The prevalence amongst mine-workers from Malawi was 4%, among those from all other countries, including South Africa, it was about 0.03%. IN response to this finding the South African Chamber of Mines stopped recruiting novices from Malawi[5] and the number of Malawians employed on the South African mines fell from 13,090 in 1988 to 2,212 in 1989.[6] These data are important for two reasons. First, they show that Malawi experienced the effects of HIV before most other countries in the region. Second, while excluding Malawian nationals from work on the mines will have seriously affected the economy of Malawi it may also, in the long run, have mitigated the impact of HIV. The spread of both HIV and TB in southern Africa is largely attributable to the system of oscillating migrant labour, introduced in the early 20th century to ensure a steady supply of cheap labour to the expanding gold mines while avoiding responsibility for the long term health of the workers.[7] This system of labour was expanded and entrenched under successive Apartheid governments and while the number of men employed in hard-rock mining has fallen substantially in the last 20 years, it still persists.

The key data on which this analysis is based are trends in the prevalence of HIV and of the coverage of anti-retroviral therapy (ART) published by UNAIDS.[2] These data suggest that the prevalence of HIV in Malawi peaked in 1998 and then continued to fall over the next five years, before ART became available in the public sector implying that there was a significant reduction in risky sexual activity at about that time.

In order to estimate the impact of early treatment on the epidemic of HIV we first consider a counterfactual in which we imagine that ART had never been made available in Malawi. We then compare this with what has happened and what will happen if the *Current Programme* of ART provision is continued and if an *Accelerated Programme* following the WHO 2013 guidelines[8] is successfully implemented between now and 2020.

Here we fit the national trend in HIV prevalence to a dynamical model to estimate current and to project future, trends in prevalence, incidence, treatment needs and deaths. We estimate the *per capita* cost of HIV/AIDS, including the cost of providing drugs, providing support to people on ART, and hospitalization and access to primary health care facilities for people who are not on ART and develop AIDS related conditions. We do not include the social and economic costs incurred when people die of AIDS so that this calculation is conservative with regards to the overall cost to society. We have not allowed for population growth or discounted the costs.

## Data

The ANC data for Malawi are sparse; in some years no data were collected, most facilities reported for only a few years, and there are few data on the early time course of the epidemic. The timing of the epidemic is important: the earlier the epidemic started the more people are likely to have advanced HIV-disease and *vice versa*. Here we use the current UNAIDS estimates of the trend in the prevalence of HIV but it would be useful to review all available prevalence data to obtain the best possible estimate of the trend in the prevalence of HIV in Malawi.

Cost data are obtained from a study carried out by the Clinton Health Access Initiative (CHAI)[9] and the key data are in Table 1.

Table 1. Costs per person per year or per test for HIV tests, anti-retroviral drugs, laboratory running costs, care and support, and clinical care for those not on ART, in each of four WHO clinical stages.

| Item | Cost (US$) |
|---|---|
| HIV test kit | 14 |
| Carrying out one test | 10 |
| Total | 24 |
| First line drugs | 66 |
| Second line drugs | 483 |
| Average | 73 |
| Personnel | 29 |
| Laboratory | 5 |
| Other | 36 |
| Laboratory running costs | 70 |
| Viral load test | 20 |
| Medications and supplements | 16 |
| Investments and other running costs | 12 |
| Care and support | 70 |
| Monitoring and support | 118 |
| Clinical stage I | 17 |
| Clinical stage II | 33 |
| Clinical stage III | 67 |
| Clinical stage IV | 134 |
| Hospitalization (average) | 63 |

An important cost relates to the cost and number of in- and out-patient days that people spend in clinics and hospitals if they are not provided with ART. Since these data are currently lacking for Malawi we use data from South Africa.[10-12] The data for South Africa suggest that the medical cost, averaged over the life-time of a person infected with HIV, is about US$800 per year. Since the *per capita* spending on health in Malawi is 8.1% of that in South Africa[13] we set the cost to US$63 per person per year, averaged over all four clinical stages, but allow it to double from each stage to the next (Table 1).

## Epidemic Model

The model is a standard dynamical model discussed in detail elsewhere.[14,15] It includes uninfected people who are susceptible to infection while infected people go through four stages of infection to death to match the known Weibull survival for people infected with HIV but not on ART.[16] To account for heterogeneity in the risk of infection we let the transmission parameter decline with the prevalence of HIV following a Gaussian curve.[14,15]

Table 2. The birth, background mortality and transition rates are fixed;[14] the other parameters are varied to optimize the fit to the trend in the prevalence of HIV shown in Figure 1.

| Parameter | Value |
|---|---|
| Adult population in 2012 (millions)[2] | 9.1 |
| Population growth rate (percent/yr)[18] | 2.7 |
| Adult mortality (percent/yr)[19] | 3.6 |
| Force of infection/yr | 0.33 |
| Prevalence at which transmission is halved (%) | 11.8 |
| Shape parameter[17] | 2.25 |
| Transition rate/yr between HIV stages off ART | 0.348 |
| Transition rate/yr between HIV stages on ART | 0.087 |

**Fitting the model**

We first fit the model to the prevalence data without including ART to get the *No ART* counterfactual against which to compare the impact of ART. We vary the prevalence of infection in 1970, which determines the timing of the epidemic; the rate of increase; and the rate at which the risk of infection declines as the prevalence increases which determines the peak value of the prevalence. The fitted values are given in Table 2. Because of the decline in the estimated prevalence from 15% in 1998 to 12% in 2007 we cannot fit the data unless we assume that there was a 50% reduction in the force of infection between 1999 and 2004.

Table 3. Model parameters for logistic functions. *Coverage* gives the asymptotic coverage; *Rate* the exponential rate of increase; *Half-max* the year when coverage reaches half the maximum value. *OR* is the odds-ratio for the number tested to the prevalence of HIV in that stage.

| | Parameter | Value |
|---|---|---|
| Behaviour change | Asymptotic reduction | 0.47 |
| | Rate *per annum* | 0.99 |
| | Half-min. year | 2001.5 |
| Passive case finding: Stage 4 | Coverage | 0.30 |
| | Rate *per annum* | 1.00 |
| | Half-max. (year) | 2006.3 |
| | OR (testing) | 2 |
| Passive case Finding: Stage 3 | Coverage | 0.30 |
| | Rate *per annum* | 1.00 |
| | Half-max. (year) | 2011.1 |
| | OR (testing) | 4 |
| Active case finding | Coverage | 0.90 |
| | Rate *per annum* | 2.00 |
| | Half-max. (year) | 2016.0 |
| | Test interval (yrs) | 1.00 |
| | OR (testing) | 10 |

To model the provision of ART in the *Current Programme* we assume that certain proportions of people in the third and fourth stages of HIV infection start ART and that coverage increases logistically. We then vary the rate and timing of the increase and the proportion of people starting treatment in each stage to match the reported ART coverage (passive case finding in Table 3).

To model the *Accelerated Programme* we further increase the proportion of people starting treatment in all WHO clinical stages to between 50% and 80% and set the rate at which this roll-out happens to 1.0/year (Table 3) and we explore the effect of different levels of coverage.

**Testing rates and costs**

We assume that a proportion of HIV-positive people are tested a certain number of times each year and a proportion of those that test positive start ART. However, many people who are not infected will also be tested and in order to determine the cost of testing we need to know the case detection rate, that is the proportion of all those tested that are infected with HIV. Under passive case-finding, people present to a health-service in Stages 3 or 4 of HIV infection. We assume that of those that present to a health service with suspicions of HIV, 50% of those in Stage 4, 25% in stage 3 and 10% of those tested under active case finding, will be infected with HIV. Since we also have to avoid the mathematical possibility that we test more people than there are in the population we set the odds ratio for the number tested to the number that are HIV-positive to 2 for stage 4, 4 for stage 3, and 10 for active case-finding.

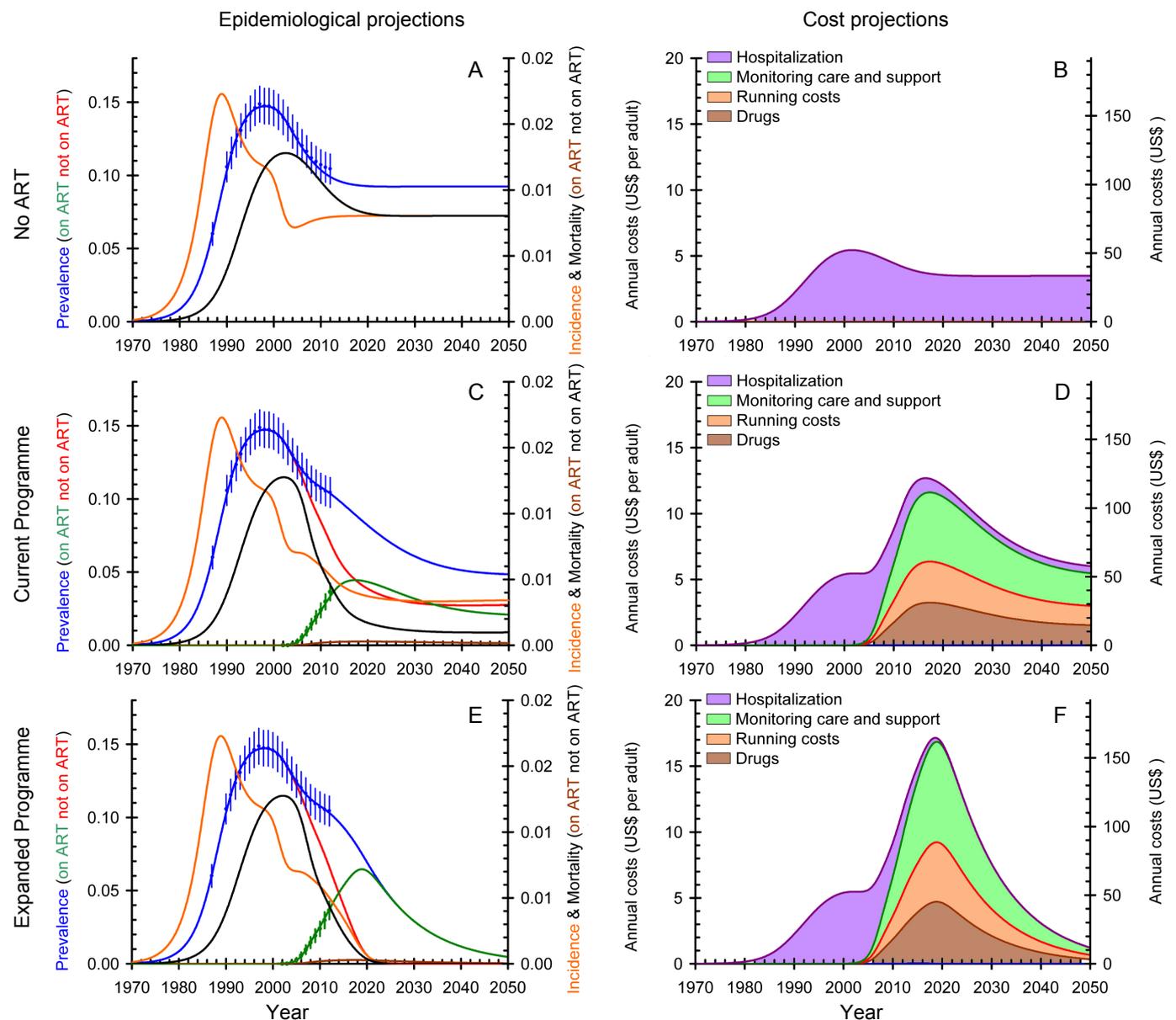

Figure 1. Top row: *No ART*. Middle row: *Current Programme*. Bottom row: *Expanded Programme*. Left*:* HIV prevalence. Blue: Data (with 95% confidence limits) and fitted line. Red: not on ART. Green: Data for those on ART and fitted line. Orange*:* HIV incidence. Black: mortality not on ART. Brown: mortality on ART. Right: Cost. Brown: Drug costs; Orange: running costs; Green: Monitoring, care and support; Purple: medical care.

## Results

The fitted data and implied trends in prevalence, incidence and mortality as well as the number on treatment are shown in Figure 1 using the parameter values in Table 2 and Table 3. Figure 1 gives the epidemiological projections on the left and the economic projections on the right. From top to bottom the graphs give the counterfactual under the *No ART* counterfactual, the *Current Programme*, and the *Expanded Programme*.

### No ART

Figure 1A gives the prevalence of HIV (blue line and data), the implied incidence (orange line) and mortality (black line). Figure 1B gives the implied costs of in-patient and out-patient care[10] (purple area). Prevalence rises rapidly, reaches a peak in 1999, declines equally rapidly for the next ten years, and then reaches a steady state at 9%. Incidence peaks in 1989 and declines as the epidemic saturates and people start to die. In 1998 the incidence starts to decline further as the force of infection falls, presumably due to changes in people's behaviour, and levels off at about 0.8% per year. Mortality rises about ten years after the incidence, reflecting the mean life-expectancy of people with HIV, and also levels off at 0.8% per year. The annual cost to the health system (Figure 1B) as a result of having to treat people with AIDS related opportunistic infections increases to US$5.4 per adult per year, or US$49M per year, in 2000 and then falls to a steady state at about US$3.5 per adult per year, or US$32M per year in total.

### Current programme

The current programme has already achieved remarkable results. The number of people on ART has increased from close to zero in 2004 to about 364 thousand in 2012 (Figure 1 C; green dots and line). This has reduced the incidence of infection in 2014 by 44% (Figure 1 A and C; orange lines) and reduced AIDS related deaths by 66% (Figure 1 A and C; black lines). The total cost of the epidemic to the country in 2014 has been increased from US$35 million to US$113 million. Up to 2014 the *Current Programme* has saved 348 thousand lives and prevented 157 thousand new infections at a cost of US$375 million (9.3 infections averted and 4.2 lives saved per US$10 thousand spent). While the cost has been considerable the savings in new infections and lives have been very significant.

### Accelerated Programme

The current (2013) guidelines of the International AIDS Society[22] and the Department of Health and Human Services (DHHS)[23] both recommend treatment for those infected with HIV, without regard to their $CD4^+$ cell count, on the grounds that this is in the best interests of the individuals concerned and has the added benefit of reducing the likelihood that they will infect their partners. The current WHO guidelines are marginally more conservative recommending immediate treatment for everyone with a $CD4^+$ cell count below 500/μL, those infected with TB, all pregnant women, all children under 5 years of age, as well as those with Hepatitis C and under these conditions an estimated 90% of people infected with HIV would be eligible to start ART as soon as they sero-convert. It would be remiss to go to great expense to avoid offering ART to the remaining 10% who would, in any event, be eligible for ART within one year.

Universal access to early treatment will reduce HIV transmission and AIDS related deaths to very low levels by 2020 (Figure 1E, orange and black lines) but there will still be a very large number of HIV positive people on ART (Figure 1E: green line) who will have to be maintained on treatment for the rest of their lives. There will be an initial increase in costs as the backlog of untreated patients is taken up, the costs will reach a peak of about US$17 per adult or US$155M in total in 2020 but will fall rapidly after that (Figure 1F) as the prevalence of infection falls. Malawi currently spends US$74 *per capita* per year on health care so that at the peak expenditure ART would take up 23% of the total health care budget.

## Conclusion

### No ART

The available data suggest that the adult prevalence of HIV in Malawi had already reached 0.9% in 1980, rose rapidly to a peak of 15% in 1998 but had declined to 12% in 2005, before ART became widely available in the public sector. This decline suggests that there was a significant reduction in sexual-risk behaviour between 1999 and 2004. If the reasons for this change can be understood and the change in behaviour reinforced this could help to strengthen prevention efforts to control the epidemic. Without the provision of ART and without further changes in behaviour, it is likely that the prevalence would have stabilized in the year 2000 at about 9.3%, the incidence and mortality at about 0.8%, so that there would have been about 74 thousand new cases and deaths every year with associated health care costs of US$32 million to which one should add the social and economic costs of letting young adults die.

### Current programme

In 2002 Malawi began an ambitious programme of ART. In 2014 an estimated 9.7% of adults, or 884 thousand people, are infected with HIV but 4.5% of adults, or 410 thousand people, are receiving ART, can be expected to live a normal life and will not infect their partners. Compared to the *No ART* counter-factual the *Current Programme* has already saved 348 thousand lives and prevented 157 thousand new infections at a cost of US$375 million (9.3 infections averted and 4.2 lives saved per US$10 thousand). If the current level of ART provision is maintained then, by 2050, this will have averted 1.6 million new infections in adults and saved 2.6 million lives at a cost of US$2.1 billion (7.6 infections averted and 12.4 lives saved per US$10 thousand). However, there will still be 255 thousand people infected with HIV, 73 thousand new infections and 9 thousand AIDS deaths each year.

### Expanded Programme

If Malawi embraces the 2013 WHO Guidelines for ART eligibility, then with 80% coverage and annual testing the *Expanded Programme* will, by 2020, further reduce the number of adults infected with HIV among adults *not* on ART to 50 thousand, the number of new infections to 6 thousand and AIDS deaths to 3 thousand. By 2025 the number of new infections and deaths will be in the hundreds per year and falling. By 2050 the *Expanded Programme* this will have averted a further 1.0 million new infections in adults and saved a further 360 thousand lives but it will also have saved Malawi US$260 million.

**Caveats**

This analysis is based on trends in the prevalence of HIV in adults derived from the *Spectrum* model.[2] These data suggest that the force of infection in Malawi fell by about 50% between 1999 and 2004. This alone will have reduced the case reproduction number from about[3] 4.4 to 2.2 so that a relatively low level of ART provision will have had a significant impact and a relatively low increase in the coverage of all infected people will end transmission. It is therefore of great importance to establish the nature and extent of this early decline in the prevalence of HIV.

**Bringing it all together**

Figure 2 compares the impact of the *Current Programme* and the potential impact of the *Expanded Programme* with the *No ART* counter factual.

The *Current Programme* has already averted 160 thousand new infections and saved 348 thousand lives. While it will continue to avert new infections and save lives the epidemic will continue indefinitely. The *Expanded Programme* will avert even more infections and save even more lives although the impact on new infections, and therefore on transmission, will be particularly important.

Between now and 2050 the cost per life saved will be close to US$816 under the current programme and US$632 under the expanded programme while the expanded programme will have saved an extra 3.2 thousand lives and US$2.3 million. Furthermore, the cost of the *Expanded Programme* will break even in about 2045 and will be cost saving after that.

**Discussion**

Malawi has experienced a severe epidemic of HIV and, given that it is a relatively poor country by regional standards, the response to both HIV and TB has been admirable. The epidemic appears to be in decline as a result of significant changes in behaviour but also the successful roll out of ART. If the response can be further strengthened then it should be possible to end AIDS in Malawi.

It is important to bear in mind that this analysis depends on the apparent decline in the prevalence of HIV before ART became widely available and the validity of this observation deserves careful confirmation. It is of interest to note that the decline in prevalence appears to have happened at about the same time as it happened in Zimbabwe where the data to support this observation are much stronger.[24] It is therefore possible that there was a significant shift in people's behaviour at about that time. It may also be relevant to note that the South African mining industry did not recruit labour from Rhodesia, now Zimbabwe, except for a brief period in the 1970s and, as noted above, they ceased to recruit men from Malawi after 1987. This exclusion from the system of oscillating migrant labour might have contributed to the decline in Malawi.

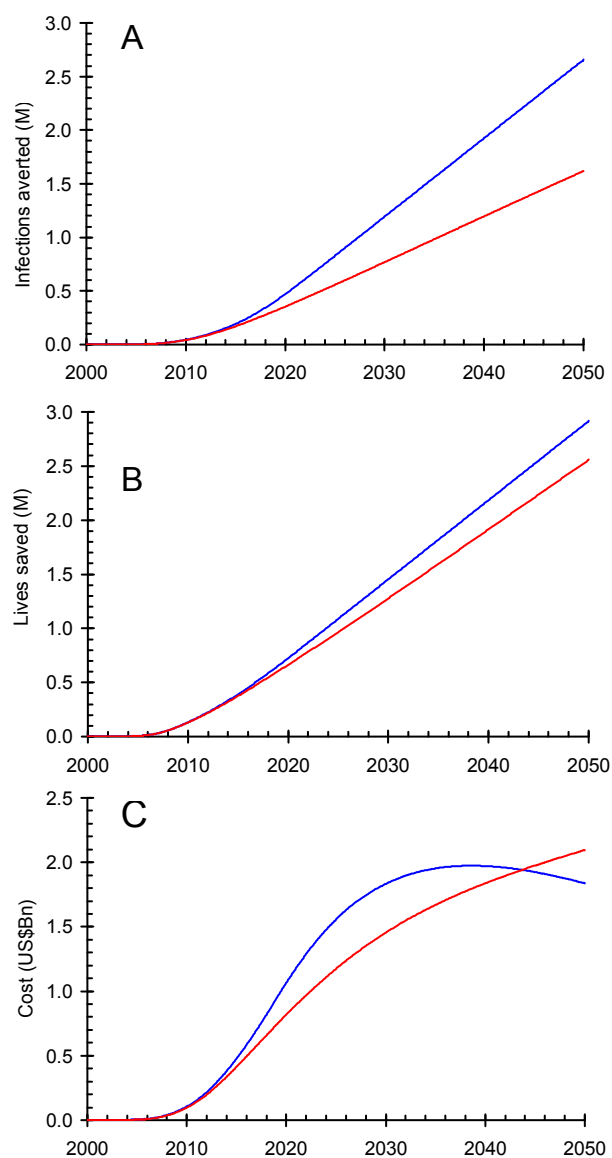

Figure 2. As compared to a counterfactual in which ART was not available the figure shows: A. Cumulative new infections averted; B. Cumulative lives saved; D. Cumulative costs. Red line: current programme; blue line expanded programme.

As in other countries in this region two critical issues are drug supply and adherence. A regular and reliable supply of drugs must be assured. Stock-outs will create anger and mistrust among infected people and poor adherence, for any reason, will lead to viral rebound, treatment failure, on-going transmission and drug resistance. The reported drop out rates of 23% at one year and 6% a year thereafter remain a cause for concern. These two considerations must be at the forefront of plans to effectively control and eventually eliminate HIV. Given the shortage of trained clinicians the programme will have to depend heavily on community mobilization and the training and support of a cadre of community health workers. However this will help to facilitate community involvement, empower local people and especially women, create jobs and stimulate local economies.

With currently available interventions universal access to early treatment is the only way to eliminate HIV in Malawi and all HIV-positive people will die of AIDS related conditions if they do not start ART; starting as soon as possible after sero-conversion will give them their best prognosis. Other methods of support and control can and should play an important supporting role.[25] To achieve high levels of compliance it will be necessary to deal with problems of stigma and discrimination and to ensure that there is strong community involvement and support for people living with HIV.

This paper focuses on treatment but continued scale-up of prevention interventions are critical to the HIV response. Condoms are highly effective if used properly and should be available to all that need them. Reducing the number of sexual partners and better control of other sexually transmitted infections are important in their own right and will contribute further to the control of HIV. Other prevention packages designed for key populations also need to be in place and can make an important contribution to stopping the epidemic of HIV; universal access to early treatment provides an ideal entry point for each of them. Finally, by developing programmes that are firmly based in local communities it will be possible to provide training and education while employing community outreach workers thereby creating jobs and stimulating local economies and ensuring sustainability.

It is still important to confirm the validity or otherwise of these results and to improve the model predictions. Here we have used the reported coverage of ART but this has never been directly measured. To do so, an appropriate sentinel surveillance system should be developed in order to carry out annual surveys on a reproducible and sufficiently large random and representative sample of adults. These surveys should include HIV testing and those that test positive for HIV should be tested for the presence of anti-retroviral drugs, their viral loads should be measured, and an incidence assay be used to estimate incidence. These would provide an important supplement to the important information already being collected and reported in the *Quarterly Reports*.

What is needed now is informed leadership and a further *Expanded Programme* of universal testing and immediate access to treatment, combined with good operational research, while monitoring the impact and implementation of the programme. The sooner that this is started the more lives and money that will be saved. Malawi has the wherewithal to stop the epidemic and the international community should support their commitment and determination to achieve an AIDS Free Generation.

## Appendix 1

The World Health Organization guidelines[26] of 2013 advise HIV-positive people to start ART if their $CD4^+$ cell count is less than $500/\mu L$, if they have TB or Hepatitis B, if they are pregnant, under the age of five years, or in a sero-discordant relationship. Data on the distribution of $CD4^+$ cell counts in HIV-negative people[27] suggest that about 80% of all those currently infected with HIV in South Africa and not on ART will have a $CD4^+$ cell count below $500/\mu L$. If we include the other groups of people that should start treatment irrespective of their $CD4^+$ cell count, then about 90% of all HIV positive people are currently eligible for ART.

If there is a need to triage people in order to treat those at greatest risk first, the sensible way to do this would be on the basis of their viral load. Individual $CD4^+$ cell counts can vary by an order of magnitude within populations,[27] the mean $CD4^+$ cell count can vary by a factor of two between populations,[27] and survival is independent of the initial $CD4^+$ cell count.[27,28] $CD4^+$ cell counts therefore have very little prognostic value except in that unfortunate circumstance when the count is very low by which time an infected person is likely to be in WHO clinical stages III or IV and in need of immediate treatment anyway. Considerable savings in time, human resources and money could be had by abandoning the use of $CD4^+$ cell counts for deciding on when to start ART. People with a high viral load, on the other hand, have a reduced life expectancy[29] and are more infectious than those with a low viral load.[30] If the availability of anti-retroviral drugs is limited it would therefore have the greatest benefit for individual patients and have the greatest impact on transmission if preference was given to people with high viral loads.[31]